\documentstyle[12pt]{article}
\begin{document}
\begin{titlepage}
\renewcommand{\thefootnote}{\fnsymbol{footnote}}

\begin{center} \LARGE
{\bf Theoretical Physics Institute}\\
{\bf University of Minnesota}
\end{center}
\begin{flushright}
TPI-MINN-94/31-T\\
UMN-TH-1312-94\\
\end{flushright}
\vspace{.3cm}
\begin{center} \LARGE
{\bf $|V_{cb}|$ from OPE Sum Rules for Heavy Flavor Transitions}
\end{center}
\begin{center}
Talk at the 27th International Conference on High Energy Physics,
Glasgow, 20-27 July 1994
\end{center}
\vspace*{.3cm}
\begin{center} {\Large
Mikhail Shifman
}
\end{center}
\vspace{0.4cm}
\begin{center}
{\it  Theoretical Physics Institute, University of Minnesota,
Minneapolis, MN 55455}\\
\vspace{.3cm}
e-mail address:
 shifman@vx.cis.umn.edu

\vspace*{.4cm}

{\Large{\bf Abstract}}
\end{center}

\vspace*{.2cm}

Recent progress in  determination of
$|V_{cb}|$ within the heavy quark expansion is reported. Both
exclusive and inclusive approaches are discussed.

\end{titlepage}

\section{Introduction}
The main topic
 today is determination of $V_{cb}$, the
CKM matrix element. We have just heard two experimental talks
devoted to measurements of this fundamental parameter.
My aim is to discuss the theoretical basis.

Two basic methods allowing one to determine $V_{cb}$ from
experimental data exist at present: exclusive and inclusive. In the
first case one studies the exclusive $B\rightarrow D^* l\nu$
decays selecting slow $D^*$'s (the so called small velocity or SV limit).
Extrapolation of the amplitude to the point of zero recoil yields
$|V_{cb}|F_{B\rightarrow D^*}$(zero recoil), where
$F_{B\rightarrow D^*}$ is an effective $B\rightarrow D^*$ transition
form factor. In the SV limit this form factor is close to unity as a
consequence of the heavy quark symmetry
\cite{Nussinov,Voloshin}; deviations from unity are quadratic in the
inverse heavy quark mass, $1/m_{b,c}^2$ \cite{Voloshin,Luke}. The
task of the theorists is to calculate these deviations.

In the inclusive approach one deals with the total semileptonic
decay rate of $B$ mesons which is proportional to $|V_{cb}|^2
m_b^5$ times a function of $m_c/m_b$. The main question is what
the quark masses actually are. The task of the theorists is to answer
this question.

So far, theoretical uncertainties quoted in the talks devoted to
determination of $|V_{cb}|$ dominate all other error bars. Reducing
them to a level significantly lower than the experimental
uncertainties is a major challenge. I am going to report today
on recent progress in this direction \cite{Bigi,Shifman}. The basic
theoretical tool is a systematic QCD-based expansion of relevant
transition operators in the inverse heavy quark masses developed
in the eighties and the very beginning of the nineties.

\section{Exclusive method}

Let me first briefly explain how one can predict deviation of
$F_{B\rightarrow D^*}$ at zero recoil from unity. To this end we
derive a sum rule for the transitions $B\rightarrow D^*$ and
$B\rightarrow$ vector
excitations generated by the axial-vector
current, $A_\mu =\bar b\gamma_\mu\gamma_5c$. If the
momentum carried by the lepton pair is denoted by $q$, the zero
recoil
point is achieved if $\vec q =0$. To obtain the sum
rule we consider the $T$ product
\begin{equation}
h_{\mu\nu} = i\int d^4x
{\rm e}^{-iqx}\frac{1}{2M_{B}}
\langle B |T\{ A_\mu^\dagger (x) A_\nu (0)\}|B\rangle
\label{1}
\end{equation}
where the hadronic tensor $h_{\mu\nu}$ can be systematically
expanded
in $\Lambda_{\rm QCD}/m_{b,c}$. For our purposes
it is sufficient to keep the terms quadratic in this parameter
and to consider only one out of five possible kinematical structures,
namely $h_1$, the only structure surviving for
the spatial components of the axial-vector current, see e.g.
\cite{Chay,Koyrakh}.

Next we use the standard technology of the sum rule approach.
Let us define
\begin{equation}
\epsilon = M_B-M_{D^*}- q_0 =\Delta M - q_0 \, .
\label{2}
\end{equation}
If $\epsilon$ is positive we sit right on the cut. The imaginary part
of the amplitude (\ref{1}) is the sum of the form factors squared
(taken at zero recoil). The sum runs over all possible intermediate
states,
$D^*$ and excitations.  We want to know  the first term in the sum,
$|F_{B\rightarrow D^*}|^2$. Alas, the
present-day QCD does not allow us to make
calculations directly in this domain.

On the other hand, if $\epsilon$ is negative we are below the cut, in
the Euclidean domain. Here the amplitude (\ref{1}) can be calculated
as an expansion in $1/m_{b,c}$
provided that
$
|\epsilon |\gg \Lambda_{\rm QCD} .
$
To get a well-defined expansion in $1/m_{b,c}$ we must
simultaneously  assume
that
$
\epsilon\ll m_{b,c}
$.

The non-perturbative corrections we are interested in are due to the
fact that both, the $c$ quark propagator connecting the points
$0$ and $x$ in Eq. (\ref{1}) and the external $b$ quark lines,
are not in the empty space but are, rather, submerged into a
soft-gluon medium, a light cloud of the $B$ meson. Two parameters
characterizing the properties of this soft medium are relevant for our
analysis. A chromomagnetic parameter
\begin{equation}
\mu_G^2 =\frac{1}{2M_B}\langle B|
\bar b\,\frac{i}{2}\sigma_{\mu\nu}G^{\mu\nu}\,b|B\rangle
=\frac{-1}{2M_B}\langle B|
\bar b\,\vec\sigma\vec B\,b|B\rangle
\label{mug}
\end{equation}
measures the correlation between the spin of the $b$ quark inside
$B$ and the chromomagnetic field $\vec B$ created by the light
cloud. The
second parameter is $\mu_\pi^2 =(2M_B)^{-1}\langle B|
\bar b\,(i\vec{D})^2 \, b|B\rangle$ measuring
the average spatial momentum squared of the $b$ quark. The both
parameters are proportional to $\Lambda_{\rm QCD}^2$. That's all
we need for the leading non-perturbative term.

If the amplitude (\ref{1}) is considered in the Euclidean domain
far below the cut (i.e. $-\epsilon \gg \Lambda_{\rm QCD}$) the
distance between the points  $0$ and $x$  is short and we can
expand $h_1$ in $\Lambda_{\rm QCD}^2/m_{b,c}^2$. Actually,  the
whole amplitude contains  more information than we need; the
sum rule sought for is obtained by considering the coefficient
in front of $1/\epsilon$ in $h_1$. In this way we arrive at the
following prediction:
$$
F_{B\rightarrow D^*}^2 + \sum_{i=1,2,...}F_{B\rightarrow excit}^2=
$$
\begin{equation}
1 -\frac{1}{3}\frac{\mu_G^2}{m_c^2}
-\frac{\mu_\pi^2-\mu_G^2}{4}\left(
\frac{1}{m_c^2}+\frac{1}{m_b^2}+\frac{2}{3m_cm_b}
\right) ,
\label{SR}
\end{equation}
where the sum on the left-hand side runs over  excited states
with the appropriate quantum numbers, up to excitation energies
$\sim \epsilon$. (In other words, $\epsilon$ plays the role of the
normalization point. Higher excited states are dual to the graphs with
the
perturbative hard gluon in the intermediate state are neglected
together with
the latter).
All form factors in Eq. (\ref{SR}) are taken at the point of zero recoil.

Let us now transfer the contribution of the excited states to the right
hand side and account for  the fact
\cite{Bigi2,Voloshin2,Bigi} that $\mu_\pi^2 >\mu_G^2$.
Then we get a lower bound on the deviation
of $F_{B\rightarrow D^*}$ from unity,
\begin{equation}
\eta_A - F_{B\rightarrow D^*} >\frac{\mu_G^2}{6m_c^2}\, .
\label{bound}
\end{equation}
Here we included the perturbative one-loop correction
\cite{Voloshin} so that $1\rightarrow\eta_A$,
\begin{equation}
\eta_A=1
+\frac{\alpha_s}{\pi}\left(\frac{m_b+m_c}{m_b-
m_c}\log{\frac{m_b}{m_c}}-
\frac{8}{3}\right)\approx 0.975\, .
\label{pertur}
\end{equation}
Using the known value of $\mu_G^2$ and $m_c =$1.3 GeV (see
below) we conclude that $F_{B\rightarrow D^*}<0.94$.

Including the $\mu_\pi^2-\mu_G^2$ term and the contribution from
the excited states lowers the prediction for $F_{B\rightarrow D^*}$
making  deviation from  unity more pronounced. If $\mu_\pi^2$
is taken from the QCD sum rule calculation \cite{Ball} the estimate
of $F_{B\rightarrow D^*}$ is reduced to 0.92. As far as the excited
states are concerned a rough estimate of the $D\pi$ intermediate
state can be given \cite{Shifman} implying that
\begin{equation}
F_{B\rightarrow D^*}=0.89\pm 0.03\, .
\label{pred}
\end{equation}
The error bars here reflect only the uncertainty in the excited states.
The parameters $\mu_\pi^2$, $\mu_G^2$, $m_c$ and $\eta_A$ have
their own error bars which I can not discuss here due to time/space
limitations.

The corrections ${\cal O}(1/m_{b,c}^2)$ to the form factors at zero
recoil have been discussed previously \cite{Falk,Mannel}
within a  version of the heavy quark expansion.
In this version, instead of the excited state contribution, one
deals with certain non-local correlation functions which are
basically unknown. About the excited states we can at least say that
their contribution has  definite sign and, moreover, we have
a rough idea of its magnitude. This is not the case for the
expansion parameters appearing in \cite{Falk,Mannel}.
It is not surprising then that even the sign of deviation of
$F_{B\rightarrow D^*}$ from unity was not understood in Ref.
\cite{Falk}, and its absolute value was underestimated.

It is curious to note that the sum rule (\ref{SR}) has been
recently questioned in Ref. \cite{Neubert} whose authors observe
an infrared contribution (due to the so called renormalons)
allegedly defying the operator product expansion. The whole
situation reminds {\em perpetuum mobile} searches. Each time a
new project is put forward always a little hurdle here or there can be
found, a crucial mistake. Sure enough, this is also the case with Ref.
\cite{Neubert}. The renormalon contribution is calculated only in the
$b$ to $c$ on-shell matrix element. Two other graphs in the
amplitude (\ref{1}),
with the gluons in the intermediate state, producing the renormalon
contribution of the same order, are simply omitted.

\section{Inclusive approach}

The CKM matrix element  $|V_{cb}|$ can be alternatively
determined from
the inclusive
semileptonic width $\Gamma (B\rightarrow X_c l\nu )$. The
theoretical
expression for
the widths is well-known
in the literature including the $\alpha_s$
and the leading non-perturbative correction, and to save space
I will not quote it here. Usually people believe that the theoretical
uncertainty is rather large since the expression for
$\Gamma (B\rightarrow X_c l\nu) $ is proportional to $m_b^5$,
and even a modest uncertainty in $m_b$ is seemingly strongly
amplified due to the fifth power.

The key observation is as follows. If one carefully examines the
formula for $\Gamma (B\rightarrow X_c l\nu )$ one observes that
it depends essentially on the difference
of the quark masses, $m_b-m_c$. This is due to the fact that in a
large part of the phase space we are not far from the SV limit, and in
the SV limit $\Gamma (B\rightarrow X_c l\nu )$ depends {\em only}
on the
difference $m_b-m_c$. For the actual values of $m_{b,c}$ the
residual
dependence on
the individual quark masses is very weak.

Now,
the quark mass
difference is known to a much better accuracy
than the individual masses,
\begin{equation}
 m_b-m_c\,= \overline{M_B}- \overline{M_D}
+
\mu_\pi^2 (\frac{1}{2m_c}-\frac{1}{2m_b}) \, +...
\label{mdif}
\end{equation}
where $\overline{M_B}=(M_B +3M_{B^*})/4$ and the same for
$\overline{M_D}$.

What is suggested? One should {\em not} allow $m_c$ to change
independently;
this parameter must be tied up to $m_b$ through Eq. (\ref{mdif}).
This simple
step dramatically reduces the uncertainty in the theoretical
prediction for $\Gamma (B\rightarrow X_cl\nu)$.

For the $b$ mass normalized not far from the would-be mass shell it
is reasonable to
accept
$ m_b = 4.8\pm 0.1$ GeV.
The central value follows from the QCD sum rule analysis
of the $\Upsilon$ system \cite{Voloshin3}. To be on a safe side  the
original error bars are multiplied by a factor of 4.  The central value
of
$m_b$ above implies $m_c \approx 1.30$ GeV (see below) which
matches
very well with an independent determination of the $c$ quark
mass \cite{Novikov}.

In this way we get numerically \cite{Shifman}
\begin{equation}
|V_{cb}| =
0.0415\left(\frac{1.49\mbox{ps}}{\tau_B}\right)^{1/2}\left(
\frac{{\rm Br}_{\rm sl}(B)}{0.106}\right)^{1/2}
\label{number}
\end{equation}
where we used the central value $4.80 {\rm GeV}$ for  $m_b$
 and the value of the
strong
coupling $\alpha_s=0.22$;  the expectation value
of  $\mu_\pi^2$ is also set equal to its central value,
$\mu_\pi^2=0.54 {\rm GeV}^2$.

What theoretical error bars in Eq. (\ref{number}) are expected?
First,
 the variation of $m_b$ (or,
alternatively, $m_c$) in the range $\pm 100 {\rm MeV}$ results only
in a
$\mp
1.6\%$ relative variation of $|V_{cb}|$ if other parameters are kept
fixed!
The most sizable uncertainty arises in this approach due to
dependence
of $m_b-m_c$ on the value of
$\mu_\pi^2$. Again, to be on a safe side, we double
the original theoretical error bars \cite{Ball} in this parameter and
allow it
to vary within the  limits
$
0.35 {\rm GeV}^2 < \mu_\pi^2 <  0.8 {\rm GeV}^2
$.
This uncertainty leads to the  change in $|V_{cb}|$ of $\mp 2.8\%$. It
seems obvious  that
the interval above  overestimates the existing uncertainty in
$\mu_\pi^2$.

It is worth noting that the value of $\mu_\pi^2$ can, and will be
measured
soon via the shape
of the lepton spectrum in $b\rightarrow c l\nu$ inclusive decays
\cite{Bigi} with
theoretical accuracy of at least $0.1 {\rm GeV}^2$.

Finally there is some dependence on the value of the strong coupling.
Numerically the uncertainty constitutes about
$\pm 1\%$ when
$\alpha_s$ is
varied
between $0.2$ and $0.25$. This must and will be reduced by explicit
calculation of
the next loop correction, which is straightforward (though somewhat
tedious
in
practice).

Therefore, the above numerical estimates imply that already at
present the theoretical uncertainty in the ``inclusive" value of
$|V_{cb}|$ does not exceed $\sim\pm 5\%$ and is quite competitive
with
the existing experimental uncertainties in this quantity. It seems
possible to
further reduce this error to 4 or even 3\%
by measuring $\mu_\pi^2$ and calculating the
two-loop perturbative correction to the width.

\section{Numerical results}

Thus, from the inclusive method we get $|V_{cb}| =0.042
\pm 0.002_{\rm theor}\pm $ experimental error.  In the exclusive
method experimentalists extrapolate to the point of zero recoil and
obtain $|V_{cb}|F_{B\rightarrow D^*}$(zero recoil).
If our central value is taken as an estimate of $F_{B\rightarrow D^*}$
then, in order to get $|V_{cb}|$ from the experimental extrapolation
to zero recoil, one must multiply the experimental number by
1.1,  quite a noticeable correction. This leads to the values
of $|V_{bc}|$ from 0.039 (CLEO) to 0.043 (ARGUS) $\pm$
experimental error. The ALEPH result lies in between. Theoretical
uncertainty in $F_{B\rightarrow D^*}$ at the level of 3 to 4$\%$
is translated in the uncertainty in $|V_{bc}|$ at the level
$\pm 0.001$ to $\pm 0.002$, i.e. slightly better although comparable
to the uncertainty one obtains in the inclusive method today.

With great satisfaction I state that the both methods nicely converge
in the problem of $V_{cb}$.

This work was supported in part by DOE under the grant number
DE-FG02-94ER40823.

\end{document}